\documentclass[twoside,twocolumn,9pt]{article}
\usepackage{extsizes}
\usepackage[super,sort&compress,comma]{natbib} 
\usepackage[version=3]{mhchem}
\usepackage[left=1.5cm, right=1.5cm, top=1.785cm, bottom=2.0cm]{geometry}
\usepackage{balance}
\usepackage{mathptmx}
\usepackage{sectsty}
\usepackage{graphicx} 
\usepackage{multirow}
\usepackage{lastpage}
\usepackage[format=plain,justification=justified,singlelinecheck=false,font={stretch=1.125,small,sf},labelfont=bf,labelsep=space]{caption}
\usepackage{float}
\usepackage{fancyhdr}
\usepackage{fnpos}
\usepackage[english]{babel}
\addto{\captionsenglish}{%
  
}
\usepackage{array}
\usepackage{droidsans}
\usepackage{charter}
\usepackage[T1]{fontenc}
\usepackage[usenames,dvipsnames]{xcolor}
\usepackage{setspace}
\usepackage[compact]{titlesec}
\usepackage{hyperref}

\usepackage{epstopdf}

\definecolor{cream}{RGB}{222,217,201}

\begin{document}

\pagestyle{fancy}
\thispagestyle{plain}
\fancypagestyle{plain}{
\renewcommand{\headrulewidth}{0pt}
}

\makeFNbottom
\makeatletter
\renewcommand\LARGE{\@setfontsize\LARGE{15pt}{17}}
\renewcommand\Large{\@setfontsize\Large{12pt}{14}}
\renewcommand\large{\@setfontsize\large{10pt}{12}}
\renewcommand\footnotesize{\@setfontsize\footnotesize{7pt}{10}}
\makeatother

\renewcommand{\thefootnote}{\fnsymbol{footnote}}
\renewcommand\footnoterule{\vspace*{1pt}%
\color{cream}\hrule width 3.5in height 0.4pt \color{black}\vspace*{5pt}} 
\setcounter{secnumdepth}{5}

\makeatletter 
\renewcommand\@biblabel[1]{#1}            
\renewcommand\@makefntext[1]%
{\noindent\makebox[0pt][r]{\@thefnmark\,}#1}
\makeatother 
\renewcommand{\figurename}{\small{Fig.}~}
\sectionfont{\sffamily\Large}
\subsectionfont{\normalsize}
\subsubsectionfont{\bf}
\setstretch{1.125} 
\setlength{\skip\footins}{0.8cm}
\setlength{\footnotesep}{0.25cm}
\setlength{\jot}{10pt}
\titlespacing*{\section}{0pt}{4pt}{4pt}
\titlespacing*{\subsection}{0pt}{15pt}{1pt}

\fancyfoot{}
\fancyfoot[LO,RE]{\vspace{-7.1pt}\includegraphics[height=9pt]{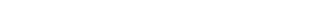}}
\fancyfoot[CO]{\vspace{-7.1pt}\hspace{13.2cm}\includegraphics{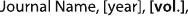}}
\fancyfoot[CE]{\vspace{-7.2pt}\hspace{-14.2cm}\includegraphics{head_foot/RF}}
\fancyfoot[RO]{\footnotesize{\sffamily{1--\pageref{LastPage} ~\textbar  \hspace{2pt}\thepage}}}
\fancyfoot[LE]{\footnotesize{\sffamily{\thepage~\textbar\hspace{3.45cm} 1--\pageref{LastPage}}}}
\fancyhead{}
\renewcommand{\headrulewidth}{0pt} 
\renewcommand{\footrulewidth}{0pt}
\setlength{\arrayrulewidth}{1pt}
\setlength{\columnsep}{6.5mm}
\setlength\bibsep{1pt}

\makeatletter 
\newlength{\figrulesep} 
\setlength{\figrulesep}{0.5\textfloatsep} 

\newcommand{\topfigrule}{\vspace*{-1pt}%
\noindent{\color{cream}\rule[-\figrulesep]{\columnwidth}{1.5pt}} }

\newcommand{\botfigrule}{\vspace*{-2pt}%
\noindent{\color{cream}\rule[\figrulesep]{\columnwidth}{1.5pt}} }

\newcommand{\dblfigrule}{\vspace*{-1pt}%
\noindent{\color{cream}\rule[-\figrulesep]{\textwidth}{1.5pt}} }

\makeatother

\twocolumn[
  \begin{@twocolumnfalse}
{\includegraphics[height=30pt]{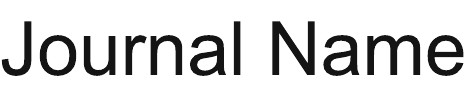}\hfill\raisebox{0pt}[0pt][0pt]{\includegraphics[height=55pt]{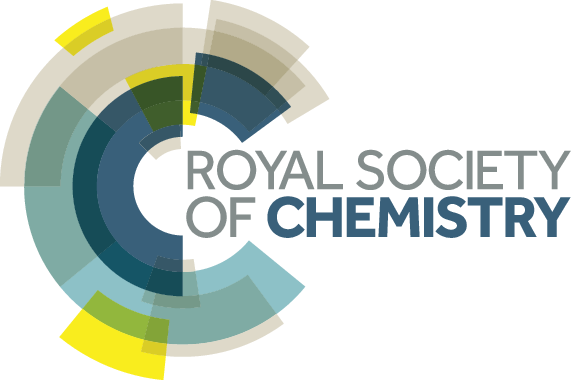}}\\[1ex]
\includegraphics[width=18.5cm]{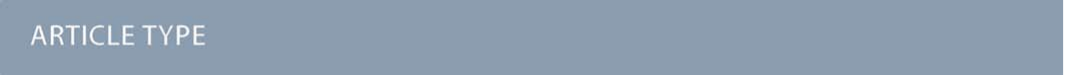}}\par
\vspace{1em}
\sffamily
\begin{tabular}{m{4.5cm} p{13.5cm} }

\includegraphics{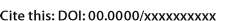} & \noindent\LARGE{\textbf{Investigating the effect of particle size distribution and complex exchange dynamics on NMR spectra of ions diffusing in disordered porous carbons through a mesoscopic model
}} \\

\vspace{0.3cm} & \vspace{0.3cm} \\

 & \noindent\large{El Hassane Lahrar,\textit{$^{a,b}$} C\'eline Merlet$^{\ast}$\textit{$^{b,c}$}} \\

\includegraphics{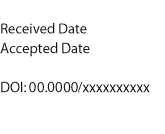} & \noindent\normalsize{Ion adsorption and dynamics in porous carbons is crucial for many technologies such as energy storage and desalination. Nuclear Magnetic Resonance (NMR) spectroscopy is a key method to investigate such systems thanks to the possibility to distinguish adsorbed (in-pore) and bulk (ex-pore) species in the spectra. However, the large variety of magnetic environments experienced by the ions adsorbed in the particles and the existence of dynamic exchange between the inside of the particles and the bulk renders the intepretation of the NMR experiments very complex. In this work, we optimise and apply a mesoscopic model to simulate NMR spectra of ions in systems where carbon particles of different sizes can be considered. We demonstrate that even for monodisperse systems, complex NMR spectra, with broad and narrow peaks, can be observed. We then show that the inclusion of polydispersity is essential to recover some experimentally observed features, such as the co-existence of peaks assigned to in-pore, exchange and bulk. Indeed, the variety of exchange rates between in-pore and ex-pore environments, present in experiments but not taken into account in analytical models, is necessary to reproduce the complexity of experimental NMR spectra.} \\

\end{tabular}

 \end{@twocolumnfalse} \vspace{0.6cm}

  ]

\renewcommand*\rmdefault{bch}\normalfont\upshape
\rmfamily
\section*{}
\vspace{-1cm}


\footnotetext{\textit{$^{a}$~Sorbonne Universit\'e, CNRS, Physicochimie des \'Electrolytes et Nanosyst\`emes Interfaciaux, F-75005 Paris, France; E-mail: el-hassane.lahrar@univ-tlse3.fr}}

\footnotetext{\textit{$^{b}$~R\'eseau sur le Stockage \'Electrochimique de l'\'Energie (RS2E), F\'ed\'eration de Recherche CNRS 3459, HUB de l'\'Energie, Rue Baudelocque, 80039 Amiens, France}}

\footnotetext{\textit{$^{c}$~CIRIMAT, Université Toulouse 3 Paul Sabatier, Toulouse INP, CNRS, Université de Toulouse, 118 Route de Narbonne, 31062 Toulouse cedex 9 - France; E-mail: celine.merlet@univ-tlse3.fr}}




\section{Introduction}

Ion dynamics in porous carbon materials play a critical role in several technological areas, such as energy storage and desalination systems.~\cite{Dou2020,Xian2021,Porada13,Luciano20} In particular, in electrochemical double layer capacitors, the energy is stored through ion adsorption at the interface between the electrode and the electrolyte. The high power density of these systems, largely related with the speed of ion adsorption/desorption, is one their main advantages. Therefore, it is crucial to have a fundamental understanding of how different ions behave under confinement in order to accurately assess the performance of porous carbons for specific uses.

Several experimental methods can be employed to investigate the adsorption and diffusion of ions in electrolytes at the interface with porous materials, such as \emph{in situ} X-ray scattering~\cite{prehal2015,prehal2017,prehal2019}, quasi-elastic neutron scattering~\cite{Dyatkin15,Dyatkin18,naresh2019}, electrochemical quartz crystal microbalance~\cite{Tsai14,Wu18,Ge24,yinguang2020,mikhael2010} and Nuclear Magnetic Resonance (NMR) spectroscopy.~\cite{john2016} These methods can provide information regarding the quantity of adsorbed ions, ion fluxes and ion dynamics to some extent.

NMR is particularly valuable for studying the properties of confined species because of its nucleus sensitivity and non-invasive nature. NMR is quantitative, which means that it can be used to accurately determine the quantities of adsorbed ions and solvent molecules, regardless of whether they are charged or neutral~\cite{john2014,john2016}. The primary factor that enables this distinction is the different chemical shift experienced by bulk (``ex-pore") and adsorbed (``in-pore") species. Indeed, in the presence of a magnetic field, the circulation of delocalised $\pi$ electrons in the sp$^2$-hybridised carbon leads to a ring-current effect~\cite{lazzert2000} which shifts the feature produced by adsorbed ions to lower frequencies in the spectrum.~\cite{robin1996,robin1999,lars2013,alex2013,michael2013}  The shift value depends on various factors such as the degree of graphitization and the pore size of the carbon~\cite{forse2014,lars2013,robert2010}. As a first approximation the shift is nucleus independent and can be estimated through Density Functional Theory (DFT) calculations on different sp$^2$-hybridized carbon structures. Such calculations of the nucleus-independent chemical shift (NICS) have provided valuable information about the measured spectra.~\cite{forse2014,xu2014,Yun-2014,Mikhail2011,Daniel2006}. 

Regarding the shape of the NMR spectra, the presence of different magnetic environments in porous carbons typically leads to a variety of linewidths for the in-pore peak. The specific linewidth depends on the similarity between these environments, which is often related to the distribution of pore sizes, as well as the motion dynamics between these different environments.~\cite{Levitt2013,Merlet2015} The change in linewidth with different electrolytes and temperatures can be used as a qualitative indicator of the dynamics of the adsorbed species. Forse~\textit{et~al.}~\cite{Forse2015} have for example shown a correlation between a decrease in temperature and an increase in the linewidth for pure ionic liquids adsorbed in an activated carbon. This increase in linewidth was attributed to the reduced mobility of the ions. Nevertheless, predicting the linewidths is challenging due to the wide variety of magnetic environments and diffusion coefficients coexisting in the complex porous carbons.~\cite{Furtado2011}. 

It is possible to realise NMR experiments more specifically focused on ion dynamics, such as pulsed-field gradient (PFG) and two-dimensional exchange spectroscopy (EXSY) experiments. Such experiments allow for the determination of diffusion coefficients and exchange rates.~\cite{todd2016,jinlei2020} Previous PFG-NMR studies have shown that a reduction in the average pore size induces a large decrease in the mobility of confined species in different porous carbons.~\cite{heink1993,dubinin1988,carmine2014,forse2017} In addition, it is known that the tortuosity of porous materials and the variation in ion concentration due to the application of an electric potential on the carbon have an effect on diffusion.~\cite{forse2017,ordintsov1998}

2D EXSY experiments are helpful to get information on the exchange of species between different environments. Griffin~\textit{et~al.}~\cite{john2014}, Deschamps~\textit{et al.}~\cite{michael2013}, and Fulik~\textit{et al.}~\cite{fulik2018} all show an in-pore/ex-pore exchange happening on the millisecond timescale in activated carbons filled with organic electrolytes. The in-pore/ex-pore exchange process was proposed to be composed of two components: a ``slow" component, which was attributed to the exchange between species that were present in the ``deep" interior region of the carbon and the bulk, and a ``fast" component, which was attributed to the exchange between species that were present at the ``surface" region of the carbon close to the bulk, and the bulk.~\cite{fulik2018}

The existence of dynamical processes occurring at different timescales suggests a possible effect of the particle size on NMR spectra. Studies on particles of few micrometers in size revealed a chemical shift averaging and a consequent change in the linewidth\cite{forse2014} which was not observed in experiments that used large particles\cite{cervini2019}. Few studies have systematically examined the impact of particle size on NMR spectra. In studies on particles of two sizes, Cervini~\textit{et~al.}\cite{cervini2020} found an exchange peak in the ex-pore environment. The slow and fast in-pore/ex-pore exchange processes for large particles have lower rate constants than for small particles, as seen by the larger peak for smaller particles. Smaller particles have shorter diffusion pathways from the inside to the exterior, resulting a faster exchange. Recently, Lyu~\textit{et~al.}~\cite{Lyu24} reported spectra of aqueous solutions in contact with a porous carbon showing three main features: an ex-pore peak, an in-pore peak and an exchange peak. The variable temperature experiments are consistent with the assignment the authors propose.

Theoretically, several analytical models have been employed to describe the exchange between two sites with different chemical shifts\cite{cavanagh2007,keeler2010}. However, their applicability is still limited to explain the sometimes complex shape of experimentally measured NMR spectra. In previous works, a mesoscopic model was developed to enables the prediction of NMR spectra for different species diffusing in a porous carbon structure.~\cite{Merlet2015} This model uses microscopic properties from molecular dynamics simulations and DFT calculations. Following its initial conception, the model has been improved to investigate \textit{in situ} NMR~\cite{anagha2021} and more recently the particle size effect.~\cite{anagha2023} In the latter case, the in-pore/ex-pore exchange can be studied by integrating a `bulk region' and a `particle region' in the system. This study revealed the importance of representing a diversity of exchange rates to explain the spectral features observed experimentally. A feature not accounted for in analytical models. Nevertheless, experimentally, the in-pore/ex-pore exchanges occurs between the bulk and particles of various sizes.~\cite{cervini2019} The inclusion of polydispersity is essential to recover specific features, such as the coexistence of ex-pore, in-pore and exchange peaks in the NMR spectra.~\cite{Lyu24} 

In this work, we introduce the possibility of having multiple particles with different sizes in the mesoscopic model to investigate the effect of the polydispersity on NMR spectra. Particles of three different sizes and two different chemical shifts for in-pore species are considered. 

\section{Methods}
\subsection{Overview of the mesoscopic lattice model}

We used the previously developed lattice-gas model~\cite{Merlet2015,anagha2023,anagha2021,belhboub2019,anagha2022} to investigate how the particle size distribution affects the NMR spectra of species diffusing through, and in and out of porous carbons. It is worth noting that this requires simulating a relatively large lattice, to include many particles, which was not possible in the implementation realised so far. The model was re-implemented using pystencils~\cite{Bauer19,Ernst23} allowing for the present study. The software, still undergoing optimisation, will be published separately. 

In the lattice model, each lattice site is either a pore or a volume of liquid for which several values need to be defined: i) the pore size (or size of the volume of liquid), ii) the quantity of ions in the pore, iii) the resonant frequency of the ions in this pore. In principle, these values can be parameterised from molecular dynamics simulations, for the quantities of ions, and from DFT, for the resonant frequencies, but in the present work, a simplified model with values consistent with experiments is chosen. 

The diffusion of electrolyte species through lattice sites is determined through an acceptance rule by which a transition from site $i$ to site $j$ follows the probability: 
\begin{equation}\label{probability}
P(i\to j) = \left\{
\begin{array}{cc}
\exp\left ( \frac{-(E_j-E_i)}{{\rm k_B}T} \right ) & \mathrm{if\ } E_j>E_i \\
1 & \mathrm{if\ } E_j\leq {E_i}
\end{array}
\right.
\end{equation}
where $E_i$ is the energy assigned to site $i$, ${\rm k_B}$ is the Boltzmann constant and $T$ is temperature of the system. A transition from a site $i$, characterized by a higher energy level, to a site $j$, characterized by a lower energy level, will always take place. The probability of the reverse transition decreases as the difference between the energies of the two sites, $E_i - E_j$ , increases.

The transition probability can be reduced by a factor $\alpha_{ij}$ to slow down diffusion. This factor is defined as $\alpha_{ij}=exp(\frac{-E_a(ij)}{{\rm k_B} T})$, where $E_a(ij)$ is the energy barrier that determines the transitions between lattice sites $i$ and $j$. While this will be used in future works to include different dynamics in the bulk and in the particles, we set $\alpha_{ij}$ to 1 for all transitions to facilitate the interpretation of the results for this first study.

\begin{figure}[ht]
\centering
  \includegraphics[scale=0.4]{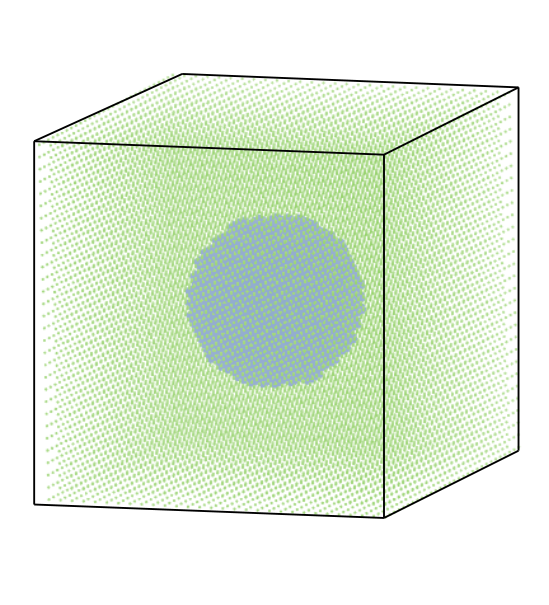}
  \caption{Scheme of the single carbon particle lattice model. A carbon particle with a radius equal to 7 lattice units (in blue) is surrounded by the bulk electrolyte (in green) for a 30$\times$30$\times$30 lattice. Note that the actual lattice size used in this work is 100$\times$100$\times$100.}
  \label{fgr:sys_mod}
\end{figure}

In the simulations conducted here, the lattice sites are assigned as bulk / ex-pore or particle / in-pore based on their positions in the system. The particles are considered spherical and are distributed randomly across the lattice. All sites that are within the radius of a given particle are particle sites. All other sites are considered to be bulk electrolyte. Figure~\ref{fgr:sys_mod} illustrates the lattice model of a single carbon particle surrounded by bulk electrolyte. 

For this first investigation, all lattice sites have the same size and contain the same quantity of ions. As a result, the ions diffuse homogeneously across the entire lattice. To determine the effect of in-pore/ex-pore exchange on the NMR spectra, different resonance frequencies are assigned to the particle and bulk sites. Since, as a first approximation, the chemical shifts observed are independent of the nucleus considered, the important quantity is the different between the bulk and in-pore chemical shifts, $\Delta\omega$. The chemical shift of the bulk sites is chosen to be 0~ppm while the chemical shift of the particle sites is either -5~ppm or -10~ppm. 

In experimental works,~\cite{Forse21} the chemical shifts observed for a range of electrolytes and carbons are between -2~ppm and -11~ppm, with most values around -5~ppm. The two values chosen here correspond to a value quite close to the average and one close to the extreme which allow us to probe the influence of the distance between the bulk and in-pore magnetic environments on the NMR spectra. It is worth noting that in real systems a distribution of environments would exist. Including such distribution is out of scope of the current work but will be explored in the future.

The NMR spectra calculations are done considering the Larmor frequency of $^{19}$F with a 300~MHz spectrometer, i.e. 282~MHz. Indeed, $^{19}$F NMR is often used to probe electrolyte ions such as BF$_4^-$ or PF$_6^-$. In the remainder of this article, we will use the words ``ions" and ``species" interchangeably as the calculations realised here do not correspond to specific carbon materials and electrolytes.

\subsection{Specific systems studied}

\begin{figure}[ht]
\centering
\includegraphics[scale=0.22]{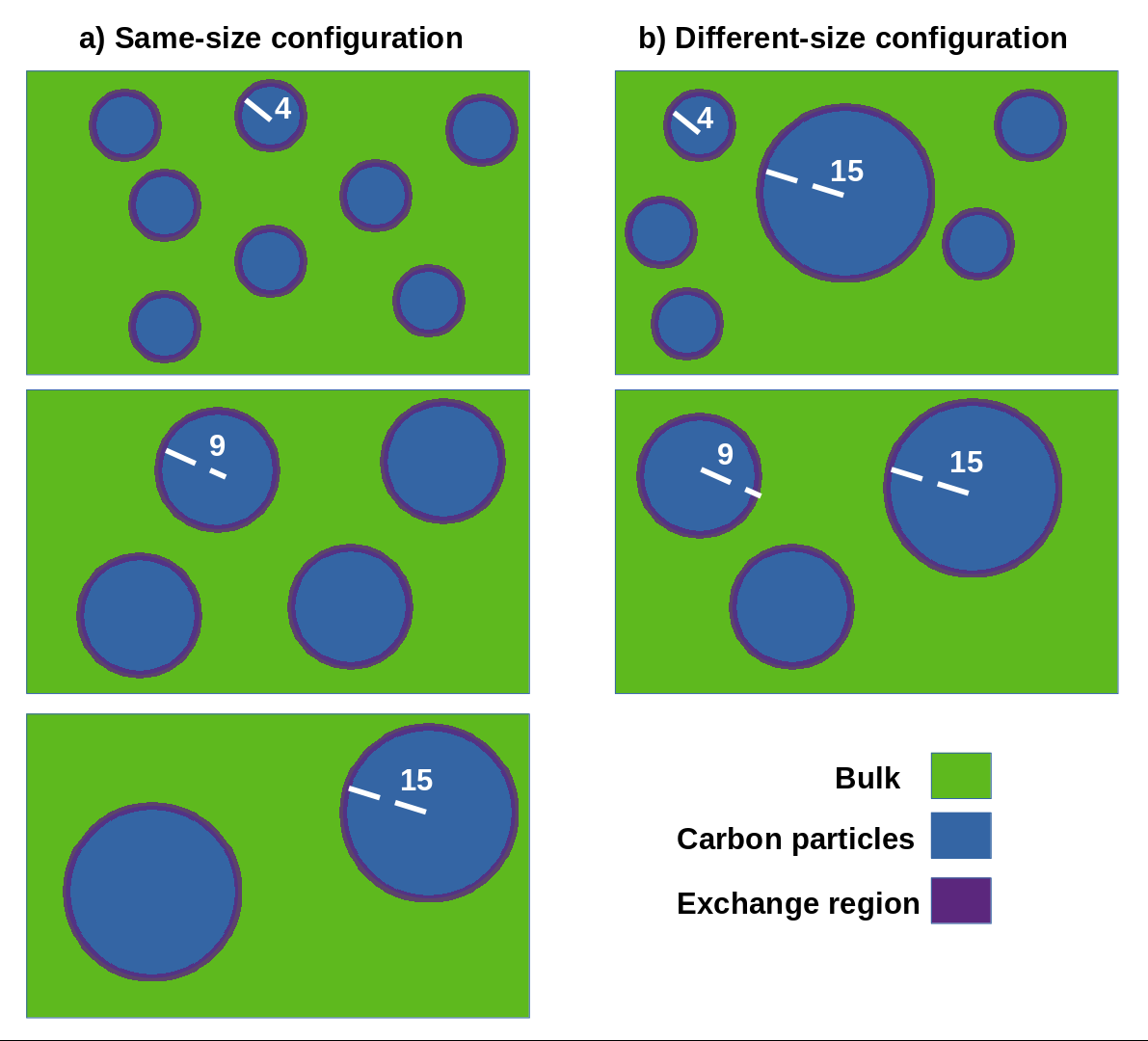}
\caption{Illustration of the systems studied for a) the same-size and b) the different-size configurations. Carbon particles are represented in blue, the bulk electrolyte in green, and in-pore/ex-pore exchange regions in purple. The real numbers of particles are given in Table~\ref{nbparticules}.}
  \label{fgr:config}
\end{figure}

In this work, for a lattice size of 100$\times$100$\times$100, two types of configurations have been considered.
\begin{itemize}
\item In the ``same-size" configuration (see Figure~\ref{fgr:config}a), the carbon particles all have the same size. The radii chosen for this configuration are 4, 9 and 15 lattice units.
\item In the ``different-size" configuration(see Figure~\ref{fgr:config}b), the particles have two different sizes. The combinations used are [4-15] and [9-15] lattice units. 
\end{itemize}
We note that in all systems the particles occupy 27\% of the total volume. The exact numbers of particles considered for all systems are provided in Table~\ref{nbparticules}.

\begin{table}[ht]
\begin{center}
  \caption{Numbers of particles considered for all systems studied.}
  \label{nbparticules}
  \begin{tabular}{|c|c|c|c|}
    \hline
    System & Size 4 & Size 9 & Size 15 \\
    \hline
    4 & 1000 & -- & -- \\
    \hline
    9 & -- & 88 & -- \\
    \hline
    15 & -- & -- & 19 \\
    \hline
    4-15 & 315 & -- & 13 \\
    \hline
    9-15 & -- & 28 & 13 \\
    \hline
  \end{tabular}
\end{center}
\end{table}

All simulations reported have been performed using 50,000 steps with a timestep of 5 $\mu$s, these settings are adequate to see a complete decay of the free induction decay signal~\cite{Merlet2015,anagha2021,belhboub2019,anagha2023}. To modify the ion dynamics, we modify the correlation time, i.e. the residence time of ions in a given site. We investigate values of $\tau$ equal to 5~ms, 2.5~ms, 1.67~ms, 1.25~ms and 1~ms.  

\section{Results and discussion}

\subsection{NMR spectra for monodisperse systems}

\begin{figure}[ht]
\centering
  \includegraphics[scale=0.38]{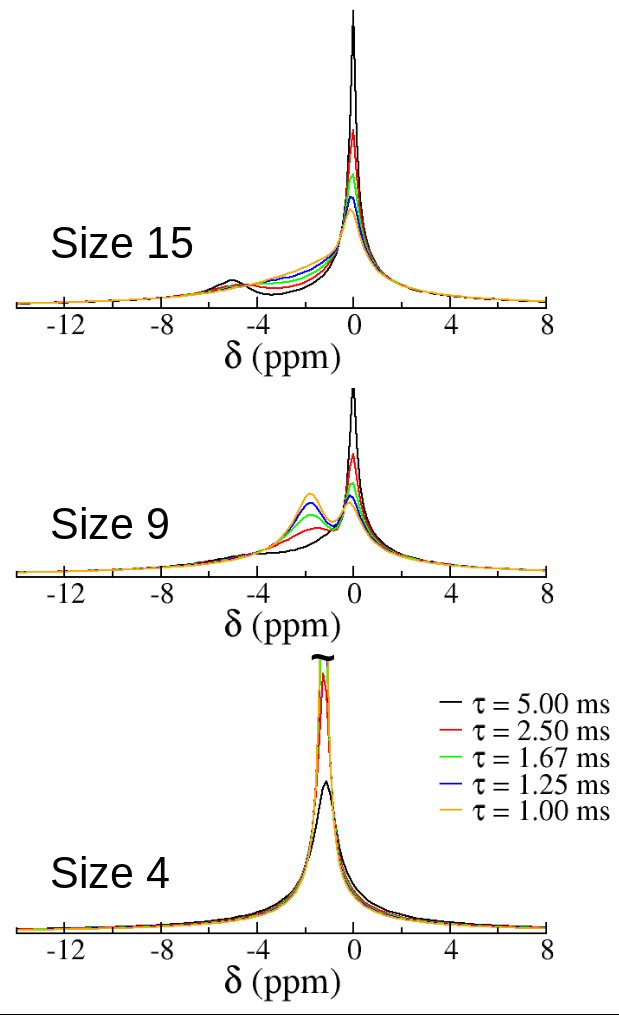}
  \caption{NMR spectra calculated for the ``same-size" configuration, for an in-pore chemical shift of -5~ppm for all particle sizes and correlation times.}
  \label{fgr:spectra-same-size-5}
\end{figure}

We first focus on monodisperse systems to explore the effect of particle size and exchange dynamics on the NMR spectra. Figure~\ref{fgr:spectra-same-size-5} shows the spectra obtained for the three different particle sizes when the chemical shift of in-pore species is -5~ppm. Peaks at this position are indeed often seen in experimental NMR spectra of various electrolytes in contact with porous carbon materials.~\cite{Forse21} 

It is very clear from Figure~\ref{fgr:spectra-same-size-5} that, as expected, both the particle size and ion dynamics have a large effect on the NMR spectra. In all spectra, either one or two peaks are observed. When there are two peaks, the position of the peak assigned to the bulk electrolyte is almost unchanged in all spectra. 

For the smallest particle size, equal to 4 lattice units, the spectra show a single peak located at approximately~-1.3~ppm. This is consistent with the fact that this system is in the fast regime, i.e. the effective exchange rate between in-pore and ex-pore species is faster than $2\pi\Delta\omega/2\sqrt2$ which means that the in-pore and ex-pore peaks are coalesced. It is worth noting that -1.3~ppm is close to the -1.35~ppm value expected from the fact that particles occupy 27\% of the lattice volume.

For the largest particle of size 15, when the species diffuse slowly enough, for correlation times larger than 2.5~ms, an in-pore peak can be clearly identified. For faster ion dynamics, the peak shifts towards larger frequencies and is partially hidden under the bulk electrolyte peak. It is worth noting that the linewidth of the bulk electrolyte peak increases when ion diffusion is faster, in agreement with the system getting closer to a full coalescence.

For the intermediate particle size, equal to 9, there are always two peaks in the spectra but the position of the peak which does not correspond to the bulk species changes dramatically between the cases where $\tau$ equals 5.00~ms and 2.50~ms. For correlation times smaller than 2.50~ms, the position of the peak at smaller frequencies is approximately -1.8~ppm. This value is closer to the -1.35~ppm, expected for a system in the fast regime, than to -5.00~ppm which would be observed in the slow exchange regime. 

These first observations with monodisperse systems already show the complexity of predicting the NMR spectra for a system with several particles. Indeed, in this precise case, while the in-pore chemical shift is always -5~ppm and the in-pore sites always occupy 27\% of the volume, the shape of the NMR spectra differ importantly. Moreover, in the case of relatively small particles, the existence of two peaks in the spectra could lead to a wrong assignment where the ``exchange" peak is thought to be an ``in-pore" peak.

\begin{figure}[ht]-
\centering
  \includegraphics[scale=0.38]{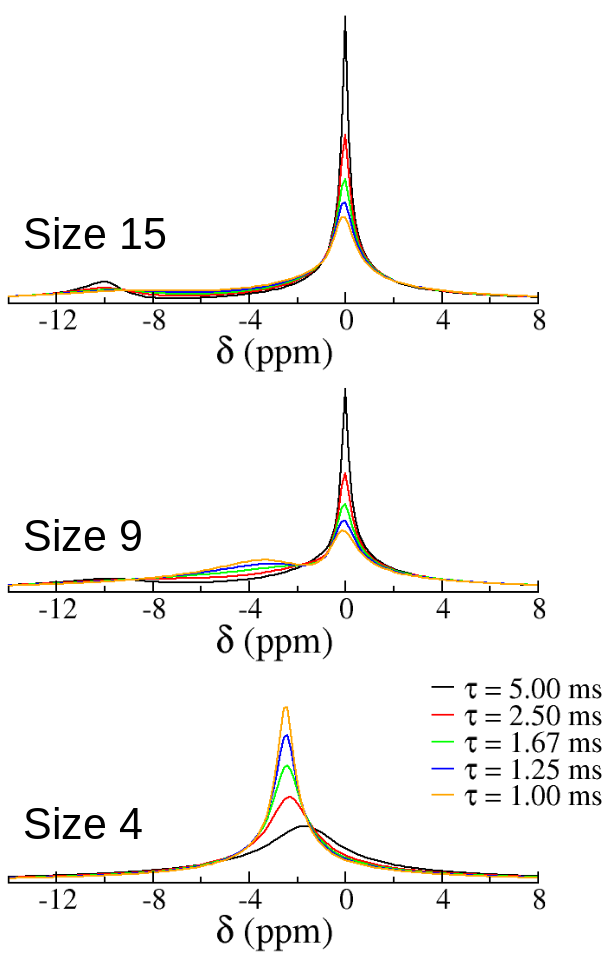}
  \caption{NMR spectra calculated for the ``same-size" configuration, for an in-pore chemical shift of -10~ppm for all particle sizes and correlation times.}
 \label{fgr:spectra-same-size-10}
\end{figure}

Figure~\ref{fgr:spectra-same-size-10} shows the spectra obtained for the three different particle sizes when the chemical shift of in-pore species is -10~ppm. Peaks at this position are not often seen in experimental NMR spectra but correspond to some of the largest values reported~\cite{Forse21} and allow us to probe the effect of $\Delta\omega$ on the spectra.

For the smallest particle size 4, the spectra again show only a single peak. However, in contrast with the case with $\Delta\omega$ equal to -5~ppm, the position and linewidth of this peak is varying more dramatically with the correlation time. This indicates that the system is closer to the coalescence point, in agreement with a larger coalescence exchange rate when $\Delta\omega$ increases by a factor of 2.

For the largest particle size 15, while the linewidth becomes quite large for correlation times smaller than 2.50~ms, the position of the peak at small frequencies is always closer to -10~ppm than to -2.7~ppm expected for a fully coalesced system. 

For the intermediate particle size 9, the two-peak spectra again show a dramatic shift of the position of the peak at lower frequencies indicating that the assignment of this peak should change from in-pore to exchange peak when the ion dynamics increase.

Overall, apart from the case of the largest particle size, the qualitative features observed between the spectra obtained with $\Delta\omega$ equal to -5~ppm and -10~ppm are not very different.
\begin{figure}[ht]
\centering
  \includegraphics[scale=0.3]{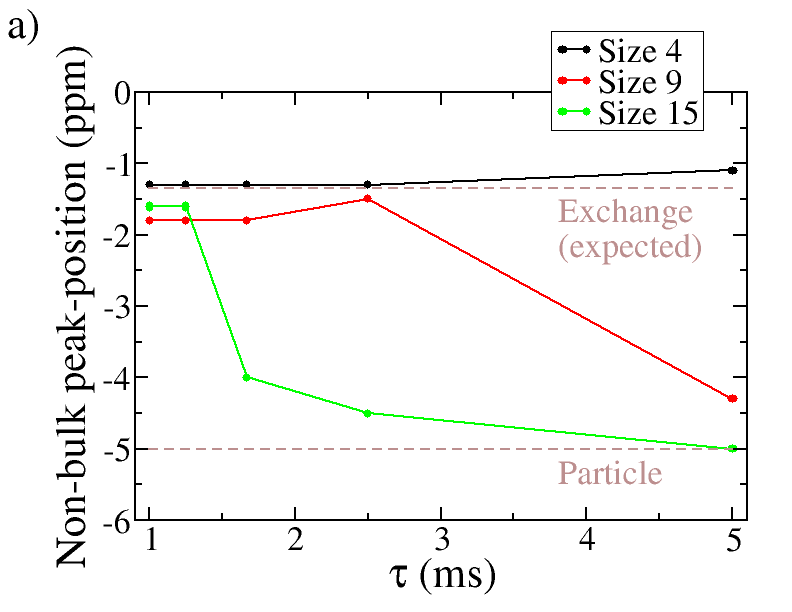}\\
  \includegraphics[scale=0.3]{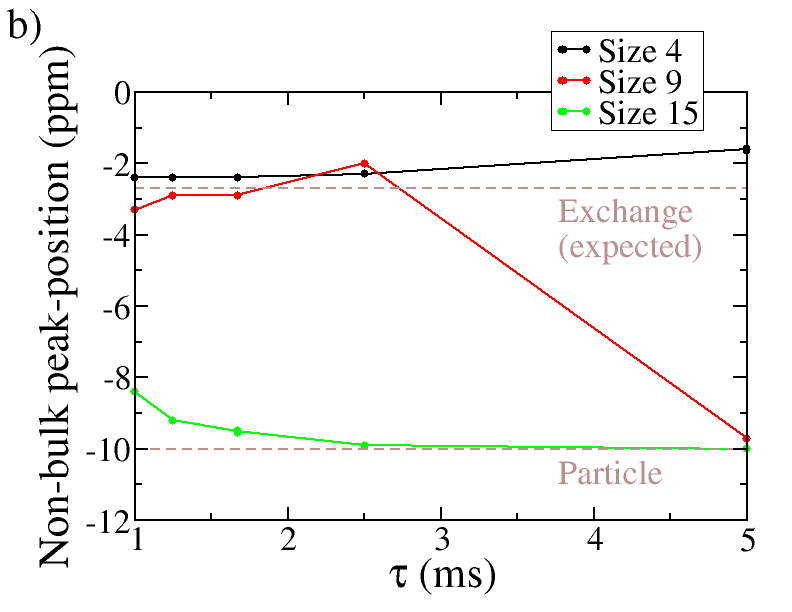}
  \caption{Peak positions of the non-bulk component of the NMR spectra for the ``same-size" systems for a particle chemical shift of a)~-5~ppm and b)~-10~ppm for all particle sizes and correlation times.}
  \label{fgr:peak-pos}
\end{figure}
Figure~\ref{fgr:peak-pos} shows the peak positions observed for the non-bulk peak. This figure confirms that the spectra show a similar behaviour for the two values of $\Delta\omega$ and that the non-bulk should sometimes be assigned to an exchange peak rather than an in-pore peak. 

Interestingly, even with monodisperse systems and without focusing on a specific carbon / electrolyte system, the resulting spectra are consistent with published data~\cite{Forse21,cervini2019,michael2013,Liu24,Lyu24} sometimes showing co-existing broad and narrow peaks.

\subsection{Effect of polydispersity on NMR spectra}

\begin{figure}[ht]
\centering
  \includegraphics[scale=0.38]{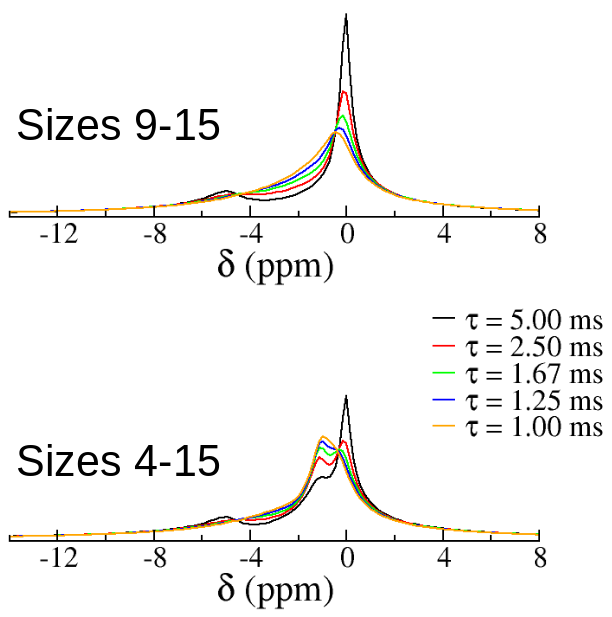}
  \caption{NMR spectra calculated for the ``different-size" configuration, for an in-pore chemical shift of -5~ppm for all particle sizes and correlation times.}
  \label{fgr:spectra-diff-size-5}
\end{figure}

\begin{figure}[ht]
\centering
  \includegraphics[scale=0.38]{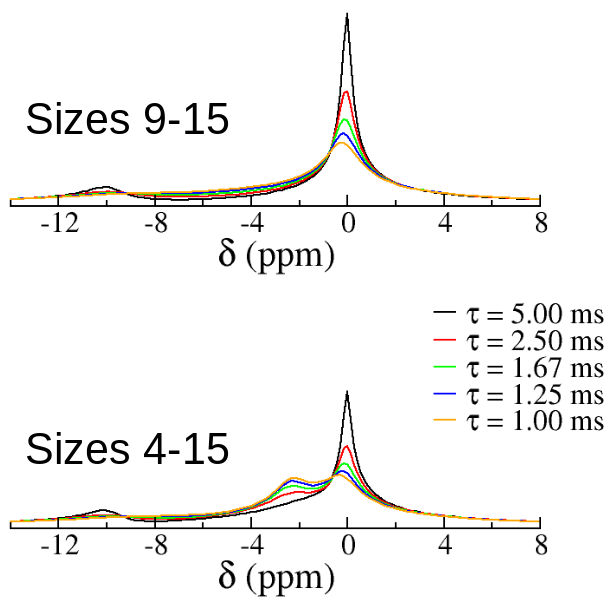}
  \caption{NMR spectra calculated for the ``different-size" configuration, for an in-pore chemical shift of -10~ppm for all particle sizes and correlation times.}
  \label{fgr:spectra-diff-size-10}
\end{figure}

We now turn to the case of polydisperse systems in which particles of two different sizes are included. Figures~\ref{fgr:spectra-diff-size-5} and~\ref{fgr:spectra-diff-size-10} show the NMR spectra obtained for the 4-15 and 9-15 systems with different $\Delta\omega$ and correlation times.
Similarly to the case of monodisperse systems, the value of $\Delta\omega$ does not change fundamentally the peaks observed in the spectra calculated.

Very interestingly, in the case of the 4-15 system, it is possible to observe three peaks in the simulated spectrum. This had been observed experimentally~\cite{Lyu24,Liu24} but, to the best of our knowledge, it is the first time that this has been simulating through a model of ion diffusion in carbon particles.
In this system, the large difference in size between the small and the large particles allow for a clear distinction between the in-pore and exchange peaks. Indeed, the effective exchange rate for small particles is much higher than the ones for large particles leading to a fast exchange and slow exchange regime co-existing.

In the case of the 9-15 system, the spectra sometimes very broad peaks which can be hard to fit experimentally. In this system, the smaller difference in size between the two types of particles leads to a different behaviour. 

It is worth noting that while the peaks observed are similar between the spectra with in-pore chemical shifts of -5~ppm and -10~ppm, the precise shape of the spectra can differ. In particular, the relative heights and linewidths of the peaks can change dramatically. This leads to spectra where broad and narrow peaks can co-exist, as is observed experimentally.~\cite{Forse21,cervini2019,michael2013,Liu24,Lyu24}

\section{Conclusions}

In this work, we have used a mesoscopic model to simulate ion diffusion in carbon particles, and between the particles and the bulk electrolyte, and investigate the effect of polydispersity on NMR spectra of such adsorbed species. 

We have shown that for monodisperse systems, NMR spectra can show one or two peaks depending on the regime, fast - intermediate - slow, in which the system is. Interestingly, there is a sharp contrast between the simplicity of the model and the complexity of the NMR spectra calculated. Indeed, while the model considered is quite simple with a single frequency assigned to in-pore lattice sites and a constant fraction of particle sites compared to bulk sites, the NMR spectra show a variety of broad and narrow peaks at different positions.

For polydisperse systems, NMR spectra show between one and three peaks, in agreement with some previously published experimental results. The co-existence of large exchange rates for small particles and small exchange rates for large particles allows for the existence of in-pore, exchange and bulk peaks in the same spectrum. This feature is, to the best of our knowledge, not observable with simpler analytical model as it results from a distribution of exchange rates inherently present in the lattice model.

This work underlines the complexity of interpreting experimental NMR spectra indicating that one has to be cautious when assigning a peak to in-pore or exchange for example. The possibility to include several particles sizes in the model opens the door for a wide range of studies where various distributions of pore sizes, particle sizes and chemical shifts can be considered. The next step will be to move towards more realistic pore size and particle size distributions. These could be obtained for example from adsorption isotherm experiments and tomographic imaging. Such avenues of research will be explored in the future.

\section*{Author Contributions}

El Hassane Lahrar: Methodology, Software, Validation, Investigation, Formal analysis, Writing – Original Draft, Writing – Review \& Editing. Céline Merlet: Methodology, Software, Validation, Writing – Original Draft, Writing – Review \& Editing, Supervision, Resources, Funding acquisition.

\section*{Conflicts of interest}

There are no conflicts to declare.

\section*{Acknowledgements}

This project has received funding from the European Union, the European High Performance Computing Joint Undertaking (JU) and countries participating in the MultiXscale project under grant agreement No 101093169. This work was granted access to the HPC resources of the CALMIP supercomputing centre under the allocations P21014. The authors acknowledge Alexander Forse, Clare Grey, John Griffin, Rudolf Weeber and Dongxun Lyu for fruitful discussions.



\balance



\providecommand*{\mcitethebibliography}{\thebibliography}
\csname @ifundefined\endcsname{endmcitethebibliography}
{\let\endmcitethebibliography\endthebibliography}{}

\end{document}